\documentstyle[12pt]{article}
\font\twlmib = cmmib10 scaled \magstep1
\def\vec#1{\mbox{\twlmib #1}}
\def\e{\mbox{e}}
\def\colvec#1#2{ \left\{ { \array{c} \!\!#1\!\! \\ \!\! #2\!\!
                           \endarray }\right\} }
\begin{document}
\title{The column vector calculus for\\
        Thermo Field Dynamics of\\
        relativistic quantum fields
\thanks{Work supported by GSI}}
\author{P.A.Henning\thanks{E-mail address: phenning@tpri6a.gsi.de}\\
        Institut f\"ur Kernphysik der TH Darmstadt and GSI\\[1mm]
        P.O.Box 110552, D-6100\thanks{after July 1st: D-64220}
        Darmstadt, Germany}
\date{May 4, 1993}
\maketitle
\begin{abstract}
A formalism is discussed which simplifies the
calculation of Feynman diagrams at finite
temperature.
\end{abstract}
\centerline{PACS No. 03.70.+k,05.30.-d,05.20.Dd,05.60.+w,05.70.Ln}
\clearpage
%%%%%%%%%%%%%%%%%%%%%%%%%%%%%%%%%%%%%%%%%%%%%%%%%%%%%%%%%%%%%%%%%%%%%
%%%%%%%%%%%%%%%%%%%%%%%%%%%%%%%%%%%%%%%%%%%%%%%%%%%%%%%%%%%%%%%%%%%%%
\section{Introduction}
Many efforts in contemporary physics are devoted to the study of
quantum systems at finite temperature or density, like e.g.
particles propagating in hot, compressed nuclear matter. While
the vacuum state of quantum field theory is invariant under
Lorentz transformations, matter states have, in general, less
symmetry. This leads to a modification of the particle spectrum
needed for the description of such a state \cite{BS75}.
For the application of perturbative (i.e. diagrammatic) methods
of quantum field theory, this modification has two main effects.

First, one has to use $2\times2$ matrix valued propagators,
as in Thermo Field Dynamics (TFD)\cite{Ubook} or the
Closed-Time-Path formalism \cite{DP91}.
Second, a consistent description requires elementary
excitations with a continuous mass spectrum, rather than
physical quasi-particles of infinite lifetime \cite{L88}.

A well defined diagrammatic expansion at finite temperature
therefore takes into account a spectral function of ``particles''
which deviates from a $\delta$-function, i.e., which is more
than a mass-shell constraint. In several
papers it was shown recently, how the propagators of such
an expansion can be diagonalized, and what the physical
meaning of the matrix structure is in the framework of
Thermo Field Dynamics \cite{YUNA92,hu92,hu92a}.

In the present work, a calculational recipe for TFD is discussed,
which was developed for non-relativistic theories \cite{CU93}.
The method is extended to relativistic quantum fields, where
it makes the calculation of Feynman diagrams for
matrix valued propagators a very simple task. To this end,
a quantum field theory with minimal coupling of fermions
and scalar uncharged bosons in thermal equilibrium
is considered
(see refs. \cite{h92fock,h93pion} for applications).

In TFD, the thermal instability of observable
states is absorbed into a Bogoliubov transformation. This
Bogliubov transformation can be written also for
interacting systems, where it defines stable, albeit non-observable
quasi-particles \cite{YUNA92,hu92,hu92a}.
At the level of the quasi-particle operators,
it can be cast into a single matrix form for the
bosonic and the fermionic sector of the model,
with a parameter $n$ that resembles the phase-space occupation
factor. While one has some freedom in parameterizing the transformation,
the most useful way to do so is
\begin{equation}\label{bdef}
 {\cal B}_{B,F}(n) =
\left(\array{cc}\left(1 \pm n\right) & -n\\
                \mp 1 & 1\endarray\right)
\;,\end{equation}
since it is linear in $n$ \cite{Ubook}.
In the fermionic sector, the $n$  are Fermi-Dirac functions,
labeled $n^+_F$ for positive
energy states and $n^-_F$ for negative energy states.
The {\em thermal quasi-particle\/} picture of TFD requires that
they depend on a continuous energy variable,
\begin{equation}
n^\pm_F(E)  = \frac{f^\pm_F(E)}{1 + f^\pm_F(E)} =
              \frac{1}{\e^{\beta( E\mp\mu_F)}+1}\;\;,\;\;
f^\pm_F(E)  =   \e^{-\beta(E \mp \mu_F)}
\;.\end{equation}
In the bosonic sector, $n$ is a Bose-Einstein distribution function,
\begin{equation}\label{nb}
n_B(E) = \frac{f_B(E)}{1-f_B(E)} = \frac{1}{\e^{\beta E}-1}
        \;\;,\;\;f_B(E) = \e^{-\beta E}
\;.\end{equation}
%%%%%%%%%%%%%%%%%%%%%%%%%%%%%%%%%%%%%%%%%%%%%%%%%%%%%%%%%%%%%%%%%%%%%%%%
\section{Full propagators in thermal equilibrium}
The full bosonic two-point Green's function, or propagator,
at finite temperature has been
derived in refs. \cite{hu92,hu92a}, as
\begin{eqnarray}
D^{(ab)}(k_0,\vec{k})&& = \int\limits_0^{\infty}\!\!dE\,
 \rho_B(E,\vec{k})\;\times\\
  &&\left(({\cal B}_B(n_B(E)))^{-1}
  \left(\!{\array{ll}
         { \frac{1}{k_0-E+i\epsilon}} & \\
    &    { \frac{1}{k_0-E-i\epsilon}}
\endarray} \right)\;
         {\cal B}_B(n_B(E))\,\tau_3\nonumber \right.\\
 - &&\left.\tau_3\,{\cal B}^T_B(n_B(E))
  \left(\!{\array{ll}
         { \frac{1}{k_0+E-i\epsilon}} & \\
    &    { \frac{1}{k_0+E+i\epsilon}}
\endarray} \right)\;
         ({\cal B}^T_B(n_B(E)))^{-1}\,\right)\nonumber
\;.\end{eqnarray}
$\rho_B(E,\vec{k})$ is a positive weight function with support only
for positive energy arguments, $\tau_3=\mbox{diag}(1,-1)$.

In the above equation, the terms
propagating particle and anti-particle states have been kept
separately. Since eqn. (\ref{nb}) can be continued to negative
energy arguments, the above propagator simplifies to
\begin{eqnarray}\label{dbk}
D^{(ab)}(k_0,\vec{k})&& = \int\limits_{-\infty}^\infty\!\!dE\,
 {\cal A}_B(E,\vec{k})\;\times\\
  &&({\cal B}_B(n_B(E)))^{-1}
  \left(\!{\array{ll}
         { \frac{1}{k_0-E+i\epsilon}} & \\
    &    { \frac{1}{k_0-E-i\epsilon}}
\endarray} \right)\;
         {\cal B}_B(n_B(E))\,\tau_3\nonumber
\;.\end{eqnarray}
${\cal A}_B(E,\vec{k})$ is the spectral function of the boson field,
\begin{equation}
{\cal A}_B(E,\vec{k}) = \rho_B(E,\vec{k})\Theta(E) -
                        \rho_B(-E,\vec{k})\Theta(-E)
\;,\end{equation}
and the limit of free particles with mass  $m$ is recovered when
\begin{equation}\label{fb}
{\cal A}_B(E,\vec{k}) \longrightarrow
\mbox{sign}(E)\,
  \delta(E^2 -\vec{k}^2 -m^2)
\;.\end{equation}
The fermion propagator is calculated in a similar way. The occurence
of a nonzero chemical potential does not hinder the above simplification,
since the occupation number parameters for positive and negative
energy states are related by
\begin{equation}
1-n^-_F(-E)= n^+_F(E)\equiv n_F(E)\;\;,\;\;
(f^-_F(-E))^{-1}=f_F^+(E)\equiv f_F(E)
\;.\end{equation}
Some elementary matrix algebra then gives, in terms of the fermionic
spectral function ${\cal A}_F(E,\vec{p})$ which has support
on the whole real energy axis,
\begin{eqnarray}\label{fbk}
&&S^{(ab)}_F(p_0,\vec{p}) =
 \int\limits_{-\infty}^\infty\!\!dE\,{\cal A}_F(E,\vec{p})\;\times\nonumber\\
&&\tau_3\, ({\cal B}_F(n_F(E)))^{-1}\;
   \left(\!{\array{ll}
         { \frac{1}{p_0-E+i\epsilon}} & \\
    &    { \frac{1}{p_0-E-i\epsilon}}
\endarray}\right)\;
 {\cal B}_F(n_F(E))
\;.\end{eqnarray}
The limit of free Dirac particles with mass $M$ is recovered when
\begin{equation}
{\cal A}_F(E,\vec{p}) \longrightarrow
(p_\mu\gamma^\mu+M)\,
\mbox{sign}(E)\,
\delta(E^2-\vec{p}^2-M^2)
\;.\end{equation}
Note, that the above equations for the full
propagators contain the full contribution of the negative energy
states.
%%%%%%%%%%%%%%%%%%%%%%%%%%%%%%%%%%%%%%%%%%%%%%%%%%%%%%%%%%%%%%%%%%%%%%%%%%
\section{Column vector notation for propagators}
To simplify the use of such matrix valued propagators,
a column vector notation was introduced
in ref. \cite{CU93}. It is obtained by rewriting
the thermal Bogoliubov matrices
defined in eqn. (\ref{bdef}) as the outer product of two column
vectors according to
\begin{equation}
\left( { \array{lr} a_1b_1 & a_1b_2 \\ a_2b_1 & a_2b_2 \endarray }
  \right) = \colvec{a_1}{a_2}  \colvec{b_1}{b_2}
\;.\end{equation}
This decomposition works only for matrices with zero determinant.
The art in applying this idea therefore lies in splitting
the propagator matrices into a sum of terms, each with determinant
zero.

For the above propagators this task is trivial: one simply isolates
the parts propagating forward and backward in time, this gives
the full fermion and boson propagators as
\begin{eqnarray}\label{fbkc}
S^{(ab)}(p_0,\vec{p})  =
   \int\limits_{-\infty}^\infty
   \!\!dE\,{\cal A}_F(E,\vec{p})\,n_F(E) \times &&\\
\left(
      {\frac{1}{p_0-E+i\epsilon}}
        \colvec{1}{1}
        \colvec{f_F^{-1}(E)}{-1}\right.
        &-&\left.
      {\frac{1}{p_0-E-i\epsilon}}
        \colvec{-1}{f_F^{-1}(E)}
        \colvec{1}{1}\right)\nonumber
\;,\end{eqnarray}
\begin{eqnarray}\label{dbkc}
D^{(ab)}(k_0,\vec{k})  =
   \int\limits_{-\infty}^\infty
   \!\!dE\,{\cal A}_B(E,\vec{k})\,n_B(E) \times &&\\
\left(
      {\frac{1}{k_0-E+i\epsilon}}
        \colvec{1}{1}
        \colvec{f_B^{-1}(E)}{1}\right.
        &-&\left.
      {\frac{1}{k_0-E-i\epsilon}}
        \colvec{1}{f_B^{-1}(E)}
        \colvec{1}{1}\right)\nonumber
\;.\end{eqnarray}
Note, that each column vector is associated with one
endpoint of a propagator line. Consequently, an
$N$-point function of the interacting theory
is described as an object composed of $N$ column vectors,
\begin{equation}
\Gamma^{(N)} = \colvec{a_1}{b_1}\colvec{a_2}{b_2}\cdots
\colvec{a_N}{b_N}
\;.\end{equation}
Now consider a typical diagram of the perturbative
expansion of an interacting
quantum field theory. In such a diagram, in general bare vertices and
irreducible vertex functions are joined by propagators.
The rule for combination of their column vectors
now is the following:\\[2mm]
{\bf Rule 1.}
If two or more propagator lines, counted by an index $1<l<L$,
meet in one $L$-point vertex, and
the vertex is still ``open'', i.e., external lines can be
attached to it, combine their column vectors to a new column
vector as
\begin{equation}
\colvec{a_1}{b_1}\times\colvec{a_2}{b_2}\times\cdots =
\colvec{\prod_l a_l}{\prod_l b_l}
\;.\end{equation}
{\bf Rule 2.}
If the vertex is ``closed'', i.e., if it is an interior vertex
and the last possible propagator
line with index $L$ is attached to it, contract it to
a scalar function
\begin{equation}
\colvec{a_L}{b_L}\bullet\colvec{\prod_l^{L-1} a_l}{\prod_l^{L-1} b_l}
  =\prod\limits_{l=1}^L a_l + \prod\limits_{l=1}^L b_l
\;.\end{equation}
Note, that in ref. \cite{CU93} a relative negative sign was used in this
rule. In the present work, this was absorbed into the definition
of the coupling constant, see below.
%%%%%%%%%%%%%%%%%%%%%%%%%%%%%%%%%%%%%%%%%%%%%%%%%%%%%%%%%%%%%%%%%%%%%%
\section{Applications of the method}
In a first step, the above rules are applied to ``one-loop''
self energy diagrams. Since the propagators of the model
contain the full spectral functions, this ``one-loop''
approximation can go substantially beyond a naive perturbation theory.
First, consider the one-loop polarization tensor,
\begin{equation}
\Pi^{(ab)}(k_0,\vec{k}) = -i \int\!\!\frac{d^4p}{(2\pi)^4}\;
  \mbox{Tr}\left[ g^{(a)} \, S^{(ab)}(p_0+k_0,\vec{p}+\vec{k})
    g^{(b)} \, S^{(ba)}(p_0,\vec{p})\right]
\;,\end{equation}
where $a$, $b$ are the thermal matrix indices taking values 1 and 2.
The coupling constants are (cf. ref. \cite{Ubook}) $g^1=-g^2=g$,
and the exchange of the indices in the second factor is achieved
by a simple transposition of the corresponding column vectors.
Hence one has
\begin{eqnarray}
\Pi^{(ab)}(k_0,\vec{k})=
-i \int\!\!\frac{d^4p}{(2\pi)^4}\,
   \int\limits_{-\infty}^\infty\!\!dEdE^\prime\;
     \mbox{Tr}\left[
    {\cal A}_F(E,\vec{p}+\vec{k})\,{\cal A}_F(E^\prime,\vec{p})
    \right]\;n_F(E)\,n_F(E^\prime) \times &&\nonumber \\
\left(
      {\frac{1}{p_0+k_0-E+i\epsilon}}
        \colvec{1}{1}
        \colvec{f_F^{-1}(E)}{-1} -
      {\frac{1}{p_0+k_0-E-i\epsilon}}
        \colvec{-1}{f_F^{-1}(E)}
        \colvec{1}{1}\right)\,\times&&\nonumber\\
\left(
      {\frac{1}{p_0-E^\prime+i\epsilon}}
        \colvec{f_F^{-1}(E^\prime)}{-1}
        \colvec{1}{1} -
      {\frac{1}{p_0-E^\prime-i\epsilon}}
        \colvec{1}{1}
        \colvec{-1}{f_F^{-1}(E^\prime)}\right)\,\times&&
\;,\end{eqnarray}
where the last $\times$ indicates that the product is closed,
i.e. the trace also applies to the thermal indices.

In performing the $p_0$-integration, out of the four
possible products in the integrand only those survive which have
different relative sign of $i\epsilon$ in their factors.
Applying rule 1, the complete polarization tensor
therefore is obtained as
\begin{eqnarray}
\Pi^{(ab)}(k_0,\vec{k})  =
-g^2\,\int\!\!\frac{d^3\vec{p}}{(2\pi)^3}\,
   \int\limits_{-\infty}^\infty\!\!dEdE^\prime\;
     \mbox{Tr}\left[
    {\cal A}_F(E,\vec{p}+\vec{k})\,{\cal A}_F(E^\prime,\vec{p})
    \right]\;n_F(E)\,n_F(E^\prime) \times &&\nonumber \\
\left(
      {\frac{1}{k_0+E^\prime-E+i\epsilon}}
        \colvec{-1}{1}
        \colvec{f_F^{-1}(E)}{-f_F^{-1}(E^\prime)} -
      {\frac{1}{k_0+E^\prime-E-i\epsilon}}
        \colvec{f_F^{-1}(E^\prime)}{-f_F^{-1}(E)}
        \colvec{-1}{1}\right)
\;.\end{eqnarray}
In the next step one uses, that this self-energy function
appears as the kernel of the Schwinger-Dyson equation for
the full boson propagator as
\begin{equation}
D(k) = D_0(k) + D_0(k)\bullet\Pi(k)\bullet D(k)
\;\end{equation}
To pick the {\em retarded\/} full propagator out of this sum,
keep only the terms with positive $i\epsilon$ in
the denominator (the retarded propagator is an analytical
function in the upper complex $k_0$-plane). Thus, the
second term on the r.h.s of this equation has, in the
retarded part, the column vector structure
\begin{eqnarray}
n_B(k_0)\colvec{1}{1}\colvec{f_B^{-1}(k_0)}{1}\bullet
  n_F(E)n_F(E^\prime)\colvec{-1}{1}
  \colvec{f_F^{-1}(E)}{f_F^{-1}(E^\prime)}\bullet\nonumber\\
n_B(k_0)\colvec{1}{1}\colvec{f_B^{-1}(k_0)}{1}& =& \nonumber\\
n_B(k_0)\colvec{1}{1}\colvec{f_B^{-1}(k_0)}{1}\;
\left(n_F(E^\prime)-n_F(E)\right)&&
\;.\end{eqnarray}
In other words, inserting an arbitrary number
of polarization functions retains {\em the same\/} column vector
structure as in eqn. (\ref{dbkc}) ! One can thus read off the above
equations the expression for the retarded polarization
function, which completely determines the retarded full boson
propagator, as
\begin{equation}\label{retpi}
\Pi^{R}(k_0,\vec{k})  =
g^2\int\!\!\frac{d^3\vec{p}}{(2\pi)^3}\,
   \int\limits_{-\infty}^\infty\!\!dEdE^\prime\;
     \mbox{Tr}\left[
    {\cal A}_F(E,\vec{p}+\vec{k})\,{\cal A}_F(E^\prime,\vec{p})
    \right]\;\left(\frac{n_F(E^\prime)-n_F(E)}{
                   k_0+E^\prime-E+i\epsilon}\right)
\;.\end{equation}
In ref. \cite{h93pion}, this equation was made the basis of
an explicit calculation of a bosonic spectral function
at finite temperature.

The second example is the Fock diagram,
\begin{equation}
\Sigma^{(ab)}(p_0,\vec{p}) = i \int\!\!\frac{d^4k}{(2\pi)^4}\;
  \mbox{Tr}\left[ g^{(a)} \, S^{(ab)}(p_0+k_0,\vec{p}+\vec{k})
    g^{(b)} \, D^{(ba)}(k_0,\vec{k})\right]
\;.\end{equation}
Following the same steps as above, i.e., contracting the
column vector products at both vertices, and performing the
$k_0$-integration such that only terms with different
sign of $i\epsilon$ survive, then gives
\begin{eqnarray}
\Sigma^{(ab)}(p_0,\vec{p})  =
g^2\,\int\!\!\frac{d^3\vec{k}}{(2\pi)^3}\,
   \int\limits_{-\infty}^\infty\!\!dEdE^\prime\;
    {\cal A}_F(E,\vec{p}+\vec{k})\,{\cal A}_B(E^\prime,\vec{k})
    \;n_F(E)\,n_B(E^\prime) \times&& \nonumber \\
\left(
      {\frac{1}{p_0+E^\prime-E+i\epsilon}}
        \colvec{1}{-1}
        \colvec{f_F^{-1}(E)}{f_B^{-1}(E^\prime)} +
      {\frac{1}{p_0+E^\prime-E-i\epsilon}}
        \colvec{f_B^{-1}(E^\prime)}{f_F^{-1}(E)}
        \colvec{1}{-1}\right)&&
\;.\end{eqnarray}
Again, put this into a Schwinger-Dyson equation
\begin{equation}
S(p) = S_0(p) + S_0(p)\bullet\Sigma(p)\bullet S(p)
\;,\end{equation}
and isolate the retarded full fermion propagator
by keeping only the terms with positive $i\epsilon$ in
the denominator. The second term on the r.h.s then has,
in the retarded part, the factors
\begin{eqnarray}
n_F(p_0)\colvec{1}{1}\colvec{f_F^{-1}(p_0)}{-1}\bullet
  n_F(E)n_B(E^\prime)\colvec{1}{-1}
  \colvec{f_F^{-1}(E)}{f_B^{-1}(E^\prime)}\bullet\nonumber\\
n_F(p_0)\colvec{1}{1}\colvec{f_F^{-1}(p_0)}{-1}& =& \nonumber\\
n_F(p_0)\colvec{1}{1}\colvec{f_F^{-1}(p_0)}{-1}\;
\left(n_B(E^\prime)-n_F(E)\right)&&
\;.\end{eqnarray}
Again, the column vector structure of the full propagator
(\ref{fbkc}) is conserved by the insertion of self energy
diagrams. From the above expression furthermore follows
the retarded self energy function as
\begin{equation}\label{retsi}
\Sigma^{R}(p_0,\vec{p})  =
g^2\,\int\!\!\frac{d^3\vec{k}}{(2\pi)^3}\,
   \int\limits_{-\infty}^\infty\!\!dEdE^\prime\;
    {\cal A}_F(E,\vec{p}+\vec{k})\,{\cal A}_B(E^\prime,\vec{k})\;
    \left(\frac{n_B(E^\prime)-n_F(E)}{
                p_0+E^\prime-E+i\epsilon}\right)
\;.\end{equation}
%%%%%%%%%%%%%%%%%%%%%%%%%%%%%%%%%%%%%%%%%%%%%%%%%%%%%%%%%%%%%%%%%%%%%%
\section{Conclusion}
In the present work, the column vector calculus of Thermo Field
Dynamics was introduced for relativistic quantum fields.
The simplicity of its application was shown in two examples:
two well known equations for retarded self energy functions
have been obtained within a few lines.

While this might not be a great gain for the generalized
one-loop examples given here, it makes higher order calculations
much simpler than with other methods. An example for
a non-relativistic vertex correction is dicussed in ref.
\cite{CU93}. The conclusion is, that the column vector
calculus is even simpler to apply than the Matsubara technique.

In  ref. \cite{h93pion}, the above eqn. (\ref{retpi})
was used for an application of TFD. There we considered a model
with two different types of fermions, i.e., nucleons and $\Delta_{33}$
resonances, coupled to a pion field, and calculated the
pionic spectral function. More sophisticated
applications are in progress.
%%%%%%%%%%%%%%%%%%%%%%%%%%%%%%%%%%%%%%%%%%%%%%%%%%%%%%%%%%%%%%%%%%%%%%

\end{document}